\documentclass{ws-ijmpb}
\usepackage{overcite}
\usepackage{graphicx}
\begin{document}

\markboth{Kh. P. Khamrakulov} {Two-soliton molecule bouncing in a
dipolar Bose-Einstein condensates under the effect of gravity}

\title{Two-soliton molecule bouncing in a dipolar Bose-Einstein
condensates under the effect of gravity}

\author{Kh. P. Khamrakulov}
\address{Physical-Technical Institute, Uzbek Academy of Sciences, 100084, Tashkent, Uzbekistan \\
khamrakulov@uzsci.net}

\maketitle

\begin{history}
\received{Day Month Year} \revised{Day Month Year} \accepted{(Day
Month Year)} \comby{(xxxxxxxxxx)}
\end{history}

\begin{abstract}
The dynamics of a two-soliton molecule bouncing on the reflecting
atomic mirror under the effect of gravity has been studied by
analytical and numerical methods. The analytical description is
based on the variational approximation. In numerical simulations,
we observe the resonance oscillations of the two-soliton's
center-of-mass position and width, induced by modulated atomic
mirror. Theoretical predictions are verified by numerical
simulations of the nonlocal Gross-Pitaevskii equation (GPE) and
qualitative agreement between them is found. Hamiltonian dynamic
system for a dipolar Bose-Einstein condensates (BECs) has been
studied.
\end{abstract}

\section{Introduction}

Bose-Einstein condensates (BECs) were first realized in 1995 on
vapors of rubidium,\cite{Wieman1} sodium,\cite{Ketterle1} and
lithium.\cite{Hulet} To observe quantum phenomenon such as
Bose-Einstein condensate, the temperature must be of order
$10^{-7}$ K. The phenomenon of Bose-Einstein condensation was
predicted in 1925 by Einstein.\cite{Einstein} Bose introduced a
method to derive the black-body spectrum for photon
gas.\cite{Bose} Einstein considered a gas of bosonic atoms, and
concluded that, below a critical temperature, a large fraction of
the atoms condenses in the lowest quantum state.\cite{Einstein}

The spin-polarized hydrogen has been a good candidate for BEC in
the late 1970s and early 1980s. Using the quantum theory of
corresponding states, Hecht\cite{Hecht} and Stwalley and
Nosanow\cite{Nosanow} concluded that spin-polarized hydrogen had
no bounds states and would remain gaseous down to zero temperature
and should be a good candidate to realize Bose-Einstein
condensation in a dilute atomic gas. Many aspects of studies of
spin-polarized hydrogen indicated that atomic systems can remain
in a metastable gaseous state close to Bose-Einstein condensation
conditions. Bose-Einstein condensation of atomic hydrogen was
realized in 1998 by Kleppner, Greytak and
collaborators.\cite{Greytak} Finally, Bose-Einstein condensation
of photons in an optical microcavity was first realized in 2010 by
Klaers and Weitz.\cite{Klaers}

An important progress has been made in the experimental cooling of
trapped polar molecules that have a large magnetic
moment.\cite{Doyle} In symmetrical molecules (such as
$\mathrm{H_{2},\,O_{2},\,N_{2}}$), the centers of gravity of the
positive and negative charges coincide in the absence of an
external electric field. Such molecules have no intrinsic dipole
moment and are called nonpolar. In asymmetrical molecules (such as
$\mathrm{CO,\, OH,\, NH_{3},\, HCl,\, CaH,\,LiCs}$), the centers
of gravity of the charges of opposite signs are displaced relative
to each other. In this case, the molecules have an intrinsic
dipole moment and are called polar.\cite{Doyle} Under the action
of an external electric field, the charges in a nonpolar molecule
become displaced relative to one another, the positive ones in the
direction of the field, the negative ones against the field. As a
result, the molecule acquires a dipole moment whose magnitude, as
shown by experiments, is proportional to the field
strength.\cite{Lahaye}

As expected, the process of polarization of a nonpolar molecule
proceeds as if the positive and negative charges of the molecule
were bound to one another by elastic forces. The action of an
external field on a polar molecule consists mainly in tending to
rotate the molecule so that its dipole moment is arranged in the
direction of the field. An external field does not virtually
affect the magnitude of a dipole moment. Consequently, a polar
molecule behaves in an external field like a rigid
dipole.\cite{Lahaye}

An experimental review\cite{Doyle} describes the rapidly
developing field of ultracold polar molecules. Molecules with
translational temperatures between 1 and 1000 mK can be easily
manipulated with electromagnetic fields and trapped.
Experimentally cold molecules were created from cold bosonic atoms
by magnetic Feshbach resonances.\cite{Doyle,Lahaye} The
experimental review\cite{Doyle} provides a rigorous description of
the methods for cooling, and trapping polar molecules. Reference
12 experimentally considered the formation of ultracold LiCs polar
molecules by a single photoassociation step beginning from laser
cooled atoms.

The University of Rostock group's paper by Mitschke \textit{et
al}.\cite{Mitschke} presents a detailed experimental observation
of temporal soliton molecules. The authors\cite{Mitschke} note
that this structure exists only in dispersion-managed fiber. They
predicted a bound state of temporal solitons in optical fibers
numerically. Mitschke's group described the first experimental
demonstration of the existence of temporal optical soliton
molecules. In recent experiment\cite{Wang} with ultracold polar
molecules, full control over the internal states of ultracold
$^{23}\mathrm{Na}\,^{87}\mathrm{Rb}$ polar molecules was obtained.
The authors used the microwave spectroscopy to control the
rotational and hyperfine states of ultracold ground state
$^{23}\mathrm{Na}\,^{87}\mathrm{Rb}$ polar molecules, which were
created by them.\cite{Wang}

Some other atoms, such as chromium, erbium, europium and
dysprosium, have a large magnetic moment of several Bohr magnetons
in their ground state, and thus experience significant magnetic
dipole-dipole interactions. Chromium,\cite{Pfau,Beaufils}
dysprosium\cite{Lu} and erbium\cite{Aikawa} were Bose-condensed.
The principal difference of chromium condensates from the alkali
atom condensates is that, chromium has a large permanent magnetic
dipole moment of $6\,\mu_{\mathrm{B}}$, and a scattering length of
about $100\,a_{0}$, where $\mu_{\mathrm{B}}$ and $a_{0}$ are the
Bohr magneton and the Bohr radius.\cite{Schmidt} This allows to
observe a perturbative effect of the dipolar interactions on the
expansion dynamics of the gas cloud.\cite{Stuhler}

In recent years, quantum bouncer problem has attracted the
attention of physicists.\cite{Baizakov,Rosanov,Akram,Golam} Gibbs
introduced the concept ``quantum bouncer''\cite{Gibbs} for the
particle. Quantum bouncer problem was extensively studied in many
pedagogical articles\cite{Langhoff,Gea-Banacloche} and original
research papers.\cite{Akram,Golam,Robinett}
Gea-Banaloche\cite{Gea-Banacloche} analytically considered quantum
bouncer problem and studied the reflection of a quantum particle
from a reflecting impenetrable atomic mirror. In a recent
work,\cite{Akram} static and dynamic properties of a weakly
interacting BEC in the quasi one-dimensional (1D) gravito-optical
surface trap were studied by analytical and numerical means.

The dynamics of bisolitonic matter-waves in a BEC that was
subjected to an expulsive harmonic potential and a gravitational
potential have been studied in Ref. 29. The authors were using a
non-isospectral scattering transform method and exact expressions
for the bright-matter–-wave bisolitons, and consequently received
solution in terms of double-lump envelopes. Some aspects of the
paper\cite{Dikande1} were experimentally
predicted.\cite{Shimizu,Mewes,Coq}

The dynamics of a matter-wave soliton on a reflecting penetrable
surface (atomic mirror) in a uniform gravitational field were
investigated in a research paper.\cite{Baizakov} In recent
review,\cite{Rosanov} the dynamics of an oscillon, videlicet, a
soliton-like cluster of the BEC on an oscillating atomic mirror in
a uniform gravitational field were studied.

The paper is organized as follows. In Sec. 2, we describe the
mathematical model and introduce the nonlocal Gross-Pitaevskii
equation (GPE). In Sec. 3, the variational approximation for the
analytical treatment of the nonlinear model is developed and its
predictions are compared to numerical simulations of the nonlocal
GPE. Section 4 is devoted to exploring the small amplitude
dynamics. In Sec. 5, we consider the dynamics of action-angle
variables and illustrate the distinctive features of the nonlinear
model. In Sec. 6, the variational approximation for the nonlocal
GPE for stationary state has been developed and applied to low
energy shape oscillations of the condensate. In concluding Sec. 7,
we summarize our results.

\section{Model Description and Governing Equation}

The atomic density at the center of a Bose-Einstein condensed
atomic cloud is usually $10^{13}-10^{15}\;\mathrm{cm^{-3}}$, which
the molecular density in air at room temperature and normal
atmospheric pressure is of order $10^{19}\;\mathrm{cm^{-3}}$. In
contrast to that in liquids and solids the density of atoms is
typically $10^{22}\;\mathrm{cm^{-3}}$. In atomic nuclei, the
density of nucleons is about $10^{38}\;\mathrm{cm^{-3}}$.

In ultracold quantum gases, a new kind of interaction via
long-range and anisotropic dipolar forces arises in addition to
short-range and isotropic contact interactions. The potential of
dipole-dipole interactions is\cite{Lahaye}
\begin{equation}
U_{dd}(r)=\frac{C_{dd}}{4\pi}\frac{1-3\cos^{2}\phi}{r^3},
\label{ddp1}
\end{equation}
where $C_{dd}=\mu_{0}\mu^2$ is the coupling constant for atoms
having a permanent magnetic dipole moment $\mu$ ($\mu_{0}$ is the
permeability of vacuum), and $C_{dd}=d^2/\varepsilon_{0}$ for
atoms having permanent electric dipole moment $d$
($\varepsilon_{0}$ is the permittivity of vacuum). The angle
$\phi$ lies between the direction of polarization and the relative
position of the atoms.

The dynamics of a dipolar BEC in the mean-field approximation at
zero temperature is governed by the three-dimensional (3D)
nonlocal GPE:\cite{Lahaye,Pitaevskii,Pethick}
\begin{equation}
i\hbar\frac{\partial\Psi}{\partial
t}=\Big[-\frac{\hbar^2}{2m}\Delta+U_{\rm{ext}}(\mathbf{r})+\frac
{4\pi
a_{s}\hbar^{2}}{m}|\Psi|^2+\int\limits_{-\infty}^{\infty}U_{dd}
(\mathbf{r}-\mathbf{r}^{\prime})|\Psi(\mathbf{r}^{\prime},t)|^2
d\mathbf{r}^{\prime}\Big]\Psi, \label{gpe1}
\end{equation}
where $\Psi(\mathbf{r},t)$ is the wave function of the condensate,
normalized to the number of atoms
$N=\int\limits_{-\infty}^{\infty}|\Psi|^2d\mathbf{r}$, $m$ is the
atomic mass, $a_{s}$ is the s-wave scattering length (below, we
shall be concerned with an attractive BEC for which $a_{s}<0$),
and
\begin{equation}
U_{\rm{ext}}(\mathbf{r})=\frac{m}{2}\big[\omega_{x}^2x^2+\omega_
{\perp}^2(y^2+z^2)\big] \label{extp}
\end{equation}
is the axially symmetric trapping potential.

When the transverse confinement is strong enough, one can assume
that the transverse dynamics are frozen, so that the dynamics are
effectively in 1D. In this case, the wave function may be
effectively factorized as
$\Psi(\mathbf{r},t)=\psi(x,t)\varphi(y,z)$, where
$\varphi(y,z)=\exp\big[-(y^2+z^2)/(2a_{\perp}^2)\big]/(\sqrt{\pi}
a_{\perp})$ is the ground state of the 2D harmonic oscillator in
the transverse direction, with
$a_{\perp}=\sqrt{\hbar/m\omega_{\perp}}$ being the transverse
harmonic oscillator length, $\omega_{\perp}$ is the transverse
trap frequency. By substituting factorized expression into the 3D
nonlocal GPE (\ref{gpe1}) and integration over the transverse
variables $y$ and $z$, one derives the 1D nonlocal GPE:
\begin{equation}
i\hbar\frac{\partial\psi}{\partial
t}=-\frac{\hbar^2}{2m}\frac{\partial^2\psi}{\partial
x^2}+\Big[U_{\mathrm{ext}}(x)+2a_{s}\hbar\omega_{\perp}|\psi|^2
+\int\limits_{-\infty}^{\infty}U_{dd}(|x-x^{\prime}|)|\psi
(x^{\prime},t)|^2dx^{\prime}\Big]\psi. \label{gpe2}
\end{equation}
The reduced 1D potential of the dipole-dipole interactions was
derived by Sinha and Santos.\cite{Santos} Thus, the $U_{dd}(x)$ is
of the form
\begin{equation}
U_{dd}(x)=-\frac{2\rho
d^2}{a_{\perp}^3}\Big[2|x|-\sqrt{\pi}(1+2x^2)e^{x^2}\mathrm{erfc}
(|x|)\Big], \label{ddp2}
\end{equation}
where $d$ is the dipole moment, $\rho$ is a variable that may
change between $\rho=1\:(\phi=0)$ and $\rho=-1/2\:(\phi=\pi/2)$.
It is necessary to note that although the dipole-dipole
interactions are long-ranged and divergent at the original 3D
potential, i. e. $r=0$, the reduced 1D potential $U_{dd}(x)$ is
regularized at $x=0$.

The BEC presents a giant matter-wave
packet.\cite{Ketterle2,Wieman2} Of special interest is the free
fall of BEC above Earth's surface. It is interesting to consider
the motion of a matter-wave packet in a gravitational field over
Earth's surface.\cite{Rosanov,Nesvizhevsky} In the present model,
the gravitational field acts on atoms in the vertical direction
and a horizontal atom mirror which reflects them back. The dancing
matter-wave packet corresponds to a stationary state.

We now assume the axial parabolic trap to be the gravitational
potential $mgx$ in its standard form, where $g$ is the
acceleration of free fall and reflects gives potential $U(x)$.
Rewrite Eq. (\ref{gpe2}) using the gravitational and reflecting
potentials
\begin{equation}
i\hbar\frac{\partial\psi}{\partial
t}=-\frac{\hbar^2}{2m}\frac{\partial^2\psi}{\partial
x^2}+mgx\psi+U(x)\psi+\Big[2a_{s}\hbar\omega_{\perp}|\psi|^2+
\int\limits_{-\infty}^{\infty}U_{dd}(|x-x^{\prime}|)|\psi
(x^{\prime},t)|^2dx^{\prime}\Big]\psi. \label{gpe3}
\end{equation}

For future analysis, rewrite Eq. (\ref{gpe3}) using the
dimensionless variables: $x\rightarrow x/l_{g},\,t\rightarrow
t/t_{g},\,l_{g}=(\hbar^2/(m^2g))^{1/3},\,t_{g}=(\hbar/(mg^2))^{1/3},
\,V(x)=U(x)/(mgl_{g}),\,d\rightarrow d/\sqrt{mgl_{g}a_{\perp}^3}$,
and the rescaled wave function
$\psi\rightarrow\sqrt{2|a_{s}|\omega_{\perp} t_{g}}\psi$,
\begin{equation}
i\frac{\partial\psi}{\partial
t}=-\frac{1}{2}\frac{\partial^2\psi}{\partial
x^2}+kx\psi+V(x)\psi+q|\psi|^2\psi-2\rho d^2\psi\int
\limits_{-\infty}^{\infty}R(|x-x^{\prime}|)
|\psi(x^{\prime},t)|^2dx^{\prime}, \label{gpe4}
\end{equation}
where $R(x)\sim U_{dd}(x)$ is the dimensionless nonlocal kernel
function. We introduced an additional dimensionless coefficient
$k=\sin\chi$ to account for the possibility of altering the effect
of gravity by changing the angle $\chi$ formed by the axis of the
quasi 1D waveguide and the horizontal reflecting potential. For
vertical position $\chi=\pi/2$ of the waveguide $k=1$, at smaller
angles $0<\chi\leq\pi/2$, then $0<k\leq 1$. Another additional
parameter $q$ is the dimensionless strength of attractive contact
interactions $(q=1)$.

To estimate quantities of the model, we consider the
$^{85}\mathrm{Rb}$ condensate, for which
$a_{s}=-20\;\mathrm{nm},\,l_{g}\approx1.3\;\mathrm{\mu
m},\,t_{g}=0.36\;\mathrm{ms}$. The transversal frequency of radial
confinement is $\omega_{\perp}=10^3\;\mathrm{rad/s}$. For the norm
$N=4$, the soliton contains approximately 720 atoms.
$^{7}\mathrm{Li}$ condensate with $a_{s}=-1.6\;\mathrm{nm}$ gives
$l_{g}\approx7\;\mathrm{\mu m},\,t_{g}=0.84\;\mathrm{ms},\,
\omega_{\perp}=10^{4}\;\mathrm{rad/s}$ and the soliton contains
approximately 1400 atoms.\cite{Baizakov,Rosanov}

\section{The Variational Approximation}

First of all, we develop the variational approximation for
arbitrary forms of the reflecting potential $V(x)$ and
gravitational potential.\cite{Anderson,Malomed} Equation
(\ref{gpe4}) may be written in the form
\begin{equation}
i\psi_{t}=-\frac{1}{2}\psi_{xx}+kx\psi+V(x)\psi+q|\psi|^2\psi-\gamma\psi\int
\limits_{-\infty}^{\infty}R(|x-x^{\prime}|)
|\psi(x^{\prime},t)|^2dx^{\prime}, \label{gpe5}
\end{equation}
where $\gamma=2\rho d^2$ is the strength of dipole-dipole
interactions and $\psi_{t}=\partial\psi/\partial
t,\,\psi_{x}=\partial\psi/\partial x$.

The kernel function $R(x)$ is complicated for the variational
analysis. In further calculations, we using the Gaussian response
function\cite{Santos,Assanto,Cuevas}
\begin{equation}
R(x)=\frac{1}{\sqrt{2\pi}w}\exp{\Big(-\frac{x^2}{2w^2}\Big)},
\label{ker}
\end{equation}
which is normalized to one
$\int\limits_{-\infty}^{\infty}R(x)dx=1$.

As a trial function for the two-soliton molecule, we use the first
Gauss-Hermite function
\begin{equation}
\psi(x,t)=A(x-\xi)\exp\bigg[-\frac{(x-\xi)^2}{2a^2}+i\,b
(x-\xi)^2+i\,v(x-\xi)+i\varphi\bigg], \label{ansatz1}
\end{equation}
where $A,\,a,\,\xi,\,v,\,b,\,\varphi$ are time dependent
variational parameters, representing the two-soliton's amplitude,
width, center-of-mass position, velocity, chirp and phase.

It should be noted that comparison between numerical simulation of
GPE and variational approximation used the Hermite-Gaussian and
super-sech ansatzes is consequential in Ref. 43. The authors
clearly established that the Hermite-Gaussian ansatz reproduces a
bisoliton only for specific signs of some parameters in the
Hermite-Gaussian ansatz. They also present special conditions
where its ansatz strictly has the profile of
bisoliton.\cite{Dikande2} It should also be noted that in the
classical paper of Lakoba and Kaup,\cite{Lakoba} conditions for
the Hermite-Gaussian ansatz are considered in order to obtain a
quasi-two-soliton solution in the nonlinear
Schr$\rm{\ddot{o}}$dinger equation.

The nonlocal GPE (\ref{gpe5}) can be obtained from the Lagrangian
density
\begin{eqnarray}
{\cal L} & = &
\frac{i}{2}\left(\psi\psi_{t}^{\ast}-\psi^{\ast}\psi_{t}\right)+
\frac{1}{2}|\psi_{x}|^2+(kx+V(x))|\psi|^2+\frac{q}{2}|\psi|^4 \nonumber \\
& - & \frac{\gamma}{2}|\psi|^2\int\limits_{-\infty}^{\infty}
R(|x-x^{\prime}|)|\psi(x^{\prime},t)|^2dx^{\prime}.
\label{lagdensity1}
\end{eqnarray}
We assume that the reflecting potential $V(x)$ is represented by
the Dirac delta function $V_{0}\delta(x)$, where $V_{0}$ is the
strength of delta potential barrier.

Now, substituting the Gauss-Hermite function (\ref{ansatz1}) and
Gaussian response function (\ref{ker}) into Lagrangian density
(\ref{lagdensity1}), one obtains
\begin{eqnarray}
{\cal L}_{1} & = & (x-\xi)^2\left[b_{t}(x-\xi)^2-2b(x-\xi)
\xi_{t}+v_{t}(x-\xi)-v\xi_{t}+\varphi_{t}\right] \nonumber \\
& \times & A^2e^{-(x-\xi)^2/a^2}, \\
{\cal L}_{2} & = & \left\{1-\frac{2(x-\xi)^2}{a^2}+(x-\xi)^2
\left[\frac{(x-\xi)^2}{a^4}+(2b(x-\xi)+v)^2\right]\right\}
\nonumber \\
& \times & \frac{A^2}{2}e^{-(x-\xi)^2/a^2}, \\
{\cal L}_{3} & = & [k\,x+V_{0}\delta(x)]A^2(x-\xi)^2e^{-(x-
\xi)^2/a^2}, \\
{\cal L}_{4} & = & \frac{qA^4}{2}(x-\xi)^4e^{-2(x-\xi)^2/a^2}, \\
{\cal L}_{5} & = & -\frac{\gamma A^4}{2\sqrt{2\pi}w}(x-\xi)^2
e^{-(x-\xi)^2/a^2}\int\limits_{-\infty}^{\infty}e^{-(x-
x^{\prime})^2/2w^2}(x^{\prime}-\xi)^2 \nonumber \\
& \times & e^{-(x^{\prime}-\xi)^2 /a^2}dx^{\prime}.
\label{lagdensity2}
\end{eqnarray}

To calculate the integral in ${\cal L}_5$, we make the change of
variables\cite{Abdullaev1} $z=\frac{1}{2}(x-x^{\prime}),
\,y=\frac{1}{2}(x+x^{\prime})$, consequently
$x=y+z,\,x^{\prime}=y-z,\,x^2+x^{\prime2}=2(y^2+z^2)$. The
Jacobian of the transformation is
$J=|x_{y}x^{\prime}_{z}-x_{z}x^{\prime}_{y}|=2$. Consequently,
$dxdx^{\prime}=2dydz$. Performing spatial integration, we obtain
the averaged Lagrangian terms
$L_{i}=\int\limits_{-\infty}^{\infty}{\cal L}_{i}dx$ are
calculated straightforwardly
\begin{eqnarray}
L_{1} & = &
\frac{\sqrt{\pi}}{2}A^2a^3\left(\frac{3}{2}a^2b_{t}-v\xi_{t}+
\varphi_{t}\right), \\
L_{2} & = & \frac{\sqrt{\pi}}{2}A^2a^3\left(\frac{3}{4a^2}+
3a^2b^2+\frac{v^2}{2}\right), \\
L_{3} & = &
\frac{\sqrt{\pi}}{2}A^2a^3\left(k\xi+\frac{2V_{0}\xi^2}
{\sqrt{\pi}a^3}e^{-\xi^2/a^2}\right), \\
L_{4} & = & \frac{\sqrt{\pi}}{2}A^2a^3\left(\frac{3q\,A^2a^2}
{16\sqrt{2}}\right), \\
L_{5} & = & -\frac{\gamma
A^4}{2\sqrt{2\pi}w}\int\limits_{-\infty}^{\infty} (x-\xi)^2
e^{-(x-\xi)^2/a^2}dx\int\limits_{-\infty}^{\infty}e^{-(x-
x^{\prime})^2/2w^2}(x^{\prime}-\xi)^2e^{-(x^{\prime}-\xi)^2
/a^2}dx^{\prime} \nonumber \\
& = & -\frac{\gamma
A^4}{\sqrt{2\pi}w}\Bigg\{\int\limits_{-\infty}^{\infty}
(y-\xi)^4\exp{\left[-\frac{2(y-\xi)^2}{a^2}\right]}
dy\int\limits_{-\infty}^{\infty}\exp\left[-\frac{2(a^2+w^2)}{(a\,w)^2}
z^2\right]dz \nonumber \\
& - & 2\int\limits_{-\infty}^{\infty}z^2\exp\left[-\frac{2(a^2+
w^2)}{(a\,w)^2}z^2\right]dz\int\limits_{-\infty}^{\infty}(y-\xi)^2
\exp\left[{-\frac{2(y-\xi)^2}{a^2}}\right]dy \nonumber
\label{lagrangian1}
\end{eqnarray}

\begin{eqnarray}
& + &
\int\limits_{-\infty}^{\infty}z^4\exp{\left[-\frac{2(a^2+w^2)}
{(a\,w)^2}z^2\right]}dz\int\limits_{-\infty}^{\infty}
\exp{\left[-\frac{2(y-\xi)^2}{a^2}\right]}dy\Bigg\} \nonumber \\
& = & -\frac{\sqrt{\pi/2}\;\gamma
A^4a^6}{32}\left[\frac{4w^2(a^2+w^2)+3a^4)}{(a^2+w^2)^{5/2}}\right].
\label{lagrangian2}
\end{eqnarray}
We recall that the norm of the two-soliton molecule corresponding
to the Gauss-Hermite function is $N=\sqrt{\pi}A^2a^3/2$.

Let us now go on to determine the form of the Lagrangian. The
averaged Lagrangian $L=L_{1}+L_{2}+L_{3}+L_{4}+L_{5}$ is
\begin{eqnarray}
L & = & N\bigg[\frac{3}{4a^2}+3a^2b^2-\frac{\xi_{t}^2}{2}+
\frac{3}{2}\,a^2b_{t}+\varphi_{t}+k\,\xi+\frac{2V_{0}\xi^2}
{\sqrt{\pi}a^3}\,e^{-\xi^2/a^2}+\frac{3\,q N}{8\sqrt{2\pi}a}\, \nonumber \\
& - & \frac{\gamma\,N}{8\sqrt{2\pi}}\cdot\frac{4w^2(a^2+w^2)+3a^4}
{(a^2+w^2)^{5/2}}\bigg], \label{lagrangian3}
\end{eqnarray}
where we have taken into account that the velocity is equal to the
time derivative of the center-of-mass position $v=\xi_{t}$.

Let us apply the usual procedure of the variational
approximation\cite{Anderson,Malomed} to the averaged Lagrangian.
Thus, the corresponding equations of motion for the width and
center-of-mass position of the two-soliton molecule are
\begin{eqnarray}
a_{tt} & = &
\frac{1}{a^3}+\frac{4V_{0}\xi^2}{\sqrt{\pi}a^4}\left(1-
\frac{2\xi^2}{3a^2}\right)e^{-\xi^2/a^2}+\frac{q\,N}{4\sqrt{2\pi}a^2} \nonumber \\
& - & \frac{\gamma\,N}{4\sqrt{2\pi}}\cdot\frac{a(a^4+4w^4)}
{(a^2+w^2)^{7/2}}, \label{maineq1}
\end{eqnarray}

\begin{eqnarray}
\xi_{tt} & = & -k-\frac{4V_{0}\xi}{\sqrt{\pi}a^3}\left(1-
\frac{\xi^2}{a^2}\right)e^{-\xi^2/a^2}. \label{maineq2}
\end{eqnarray}

The coupled system of Eqs. (\ref{maineq1}) and (\ref{maineq2})
represents the main result of this paper. Its fixed points
$(a_{0},\xi_{0})$ provide the stationary width of the two-soliton
molecule and its distance from the delta barrier, where the
actions of the gravity and reflecting potential $V(x)$ cancel each
other. The dynamics of small amplitude oscillations of the
two-soliton's width and center-of-mass position near the
stationary state can be described as the motion of a unit mass
particle in the anharmonic potentials $U_{1}(a)$ and $U_{2}(\xi)$
\begin{eqnarray}
a_{tt}=-\frac{\partial U_{1}(a)}{\partial a},\quad U_{1}(a) & = &
\frac{1}{2a^2}+\frac{4V_{0}\xi_{0}^2}{3\sqrt{\pi}a^3}\,
e^{-(\xi_{0}/a)^2}+\frac{q N}{4\sqrt{2\pi}a} \nonumber \\
& - & \frac{\gamma N}{12\sqrt{2\pi}}\cdot\frac{4w^2(a^2+w^2)+3a^4}
{(a^2+w^2)^{5/2}}, \label{pot1}
\end{eqnarray}
\begin{eqnarray}
\xi_{tt}=-\frac{\partial U_{2}(\xi)}{\partial\xi},\quad
U_{2}(\xi)=k\xi+\frac{2V_{0}\xi^2}{\sqrt{\pi}a_{0}^3}\,
e^{-(\xi/a_{0})^2}. \label{pot2}
\end{eqnarray}

\begin{figure}[htb]
\centerline{\includegraphics[width=6cm,height=5cm]{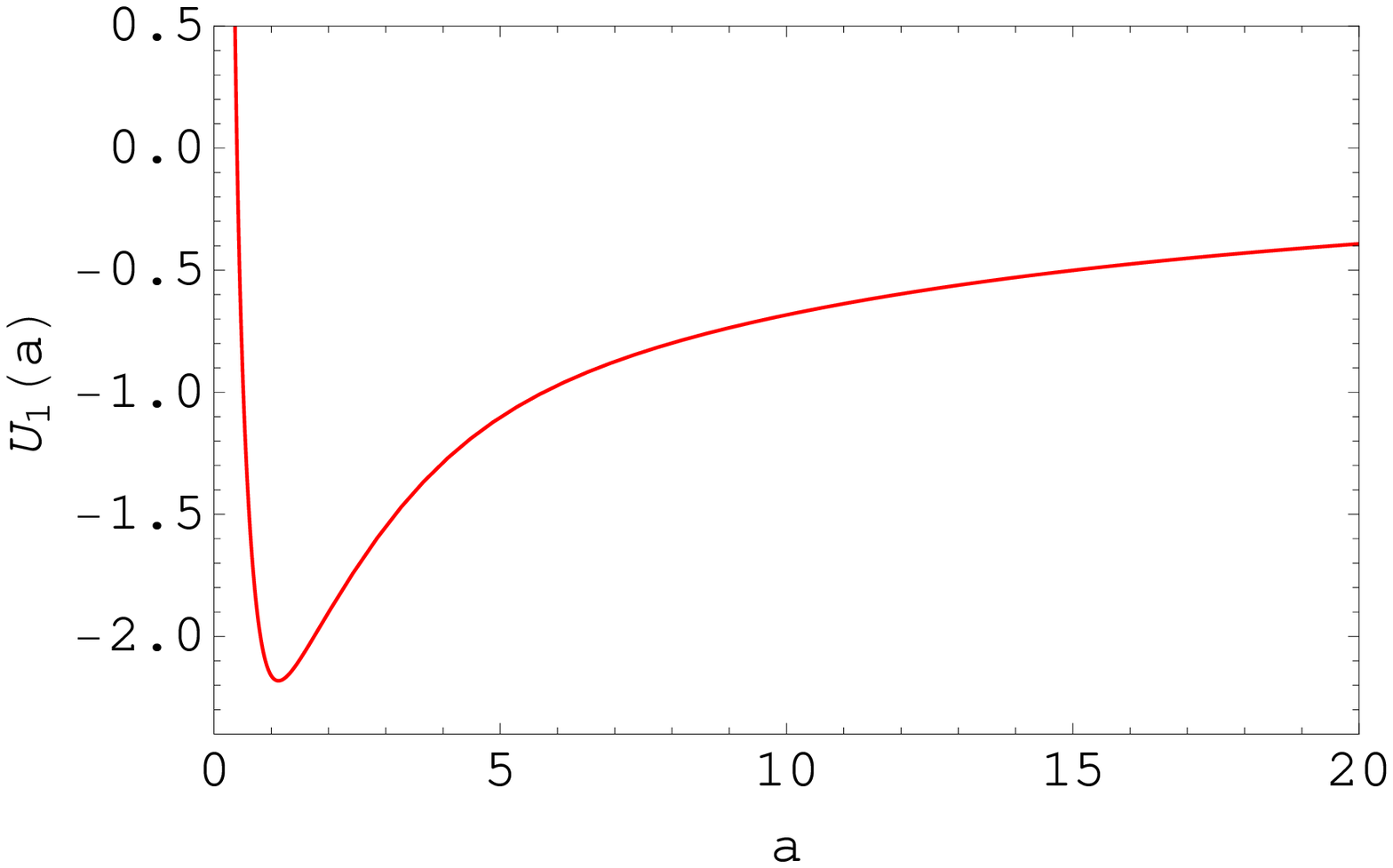}\quad
            \includegraphics[width=6cm,height=5cm]{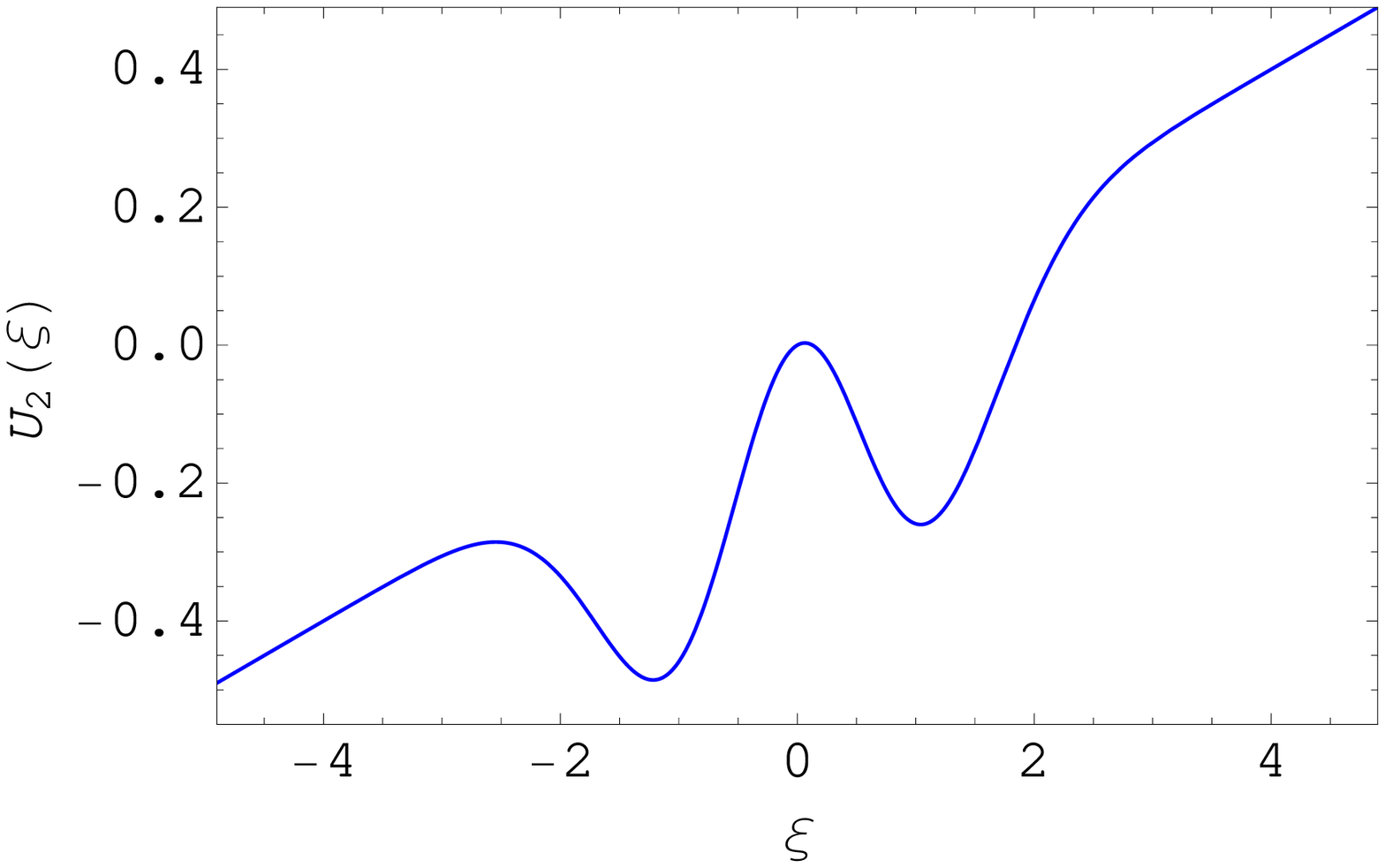}}
\caption{(Color online) The shapes of the anharmonic potentials
given by Eqs. (\ref{pot1}) and (\ref{pot2}) for attractive
interactions. Parameter values:
$V_{0}=1,\;k=0.1,\;N=4,\;q=1,\;w=5,\;d=5$ and $\gamma=20$.}
\label{fig1}
\end{figure}

\begin{figure}[htb]
\centerline{\includegraphics[width=6cm,height=5cm]{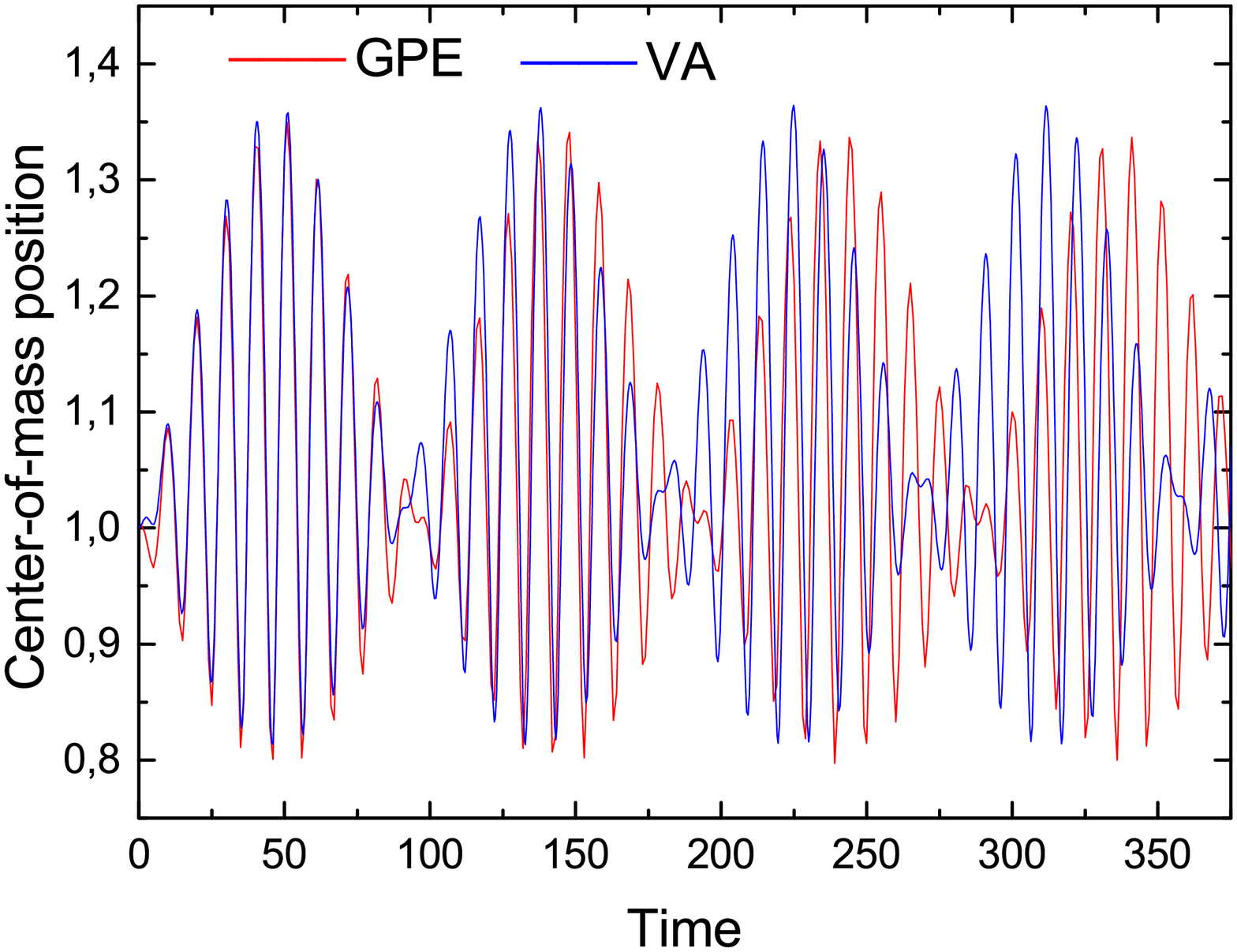}\quad
            \includegraphics[width=6cm,height=5cm]{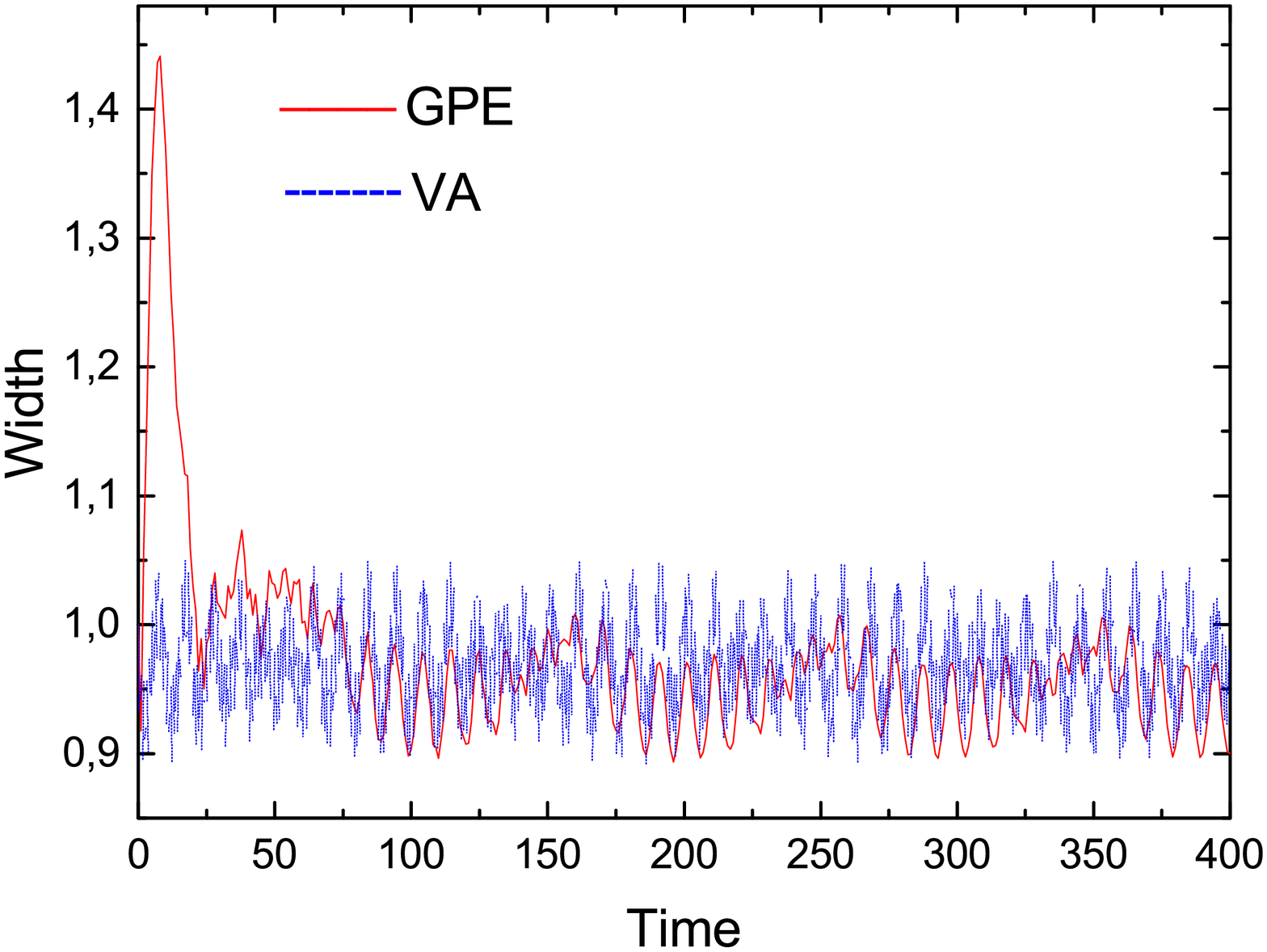}}
\caption{(Color online) Parametric excitation of a two-soliton
molecule in a dipolar BEC. Left panel: Numerical solutions of Eqs.
(\ref{maineq1}) and (\ref{maineq2}) (blue solid line) and the
governing GPE (\ref{gpe5}) (red solid line) for the center-of-mass
position of a two-soliton molecule. Right panel: Numerical
solutions of the same equations (blue dashed line) and the
governing GPE (\ref{gpe5}) (red solid line) for the width of a
two-soliton molecule. Parameters are the same as in Fig.
\ref{fig1}.} \label{fig2}
\end{figure}

Figure \ref{fig1} shows the shapes of the anharmonic potentials
$U_{1}(a)$ and $U_{2}(\xi)$ for attractive interactions. As can be
seen from this figure, the center-of-mass position of the
two-soliton molecule has two local minima. In this case, the
two-soliton molecule has a bound state. In Fig. \ref{fig2}, the
two-soliton molecule bouncing dynamics over the reflecting
potential, modelled by a delta function, are illustrated. The
result of numerical solution of the GPE is almost
indistinguishable from the prediction of variational
approximation.

Figure \ref{fig2} illustrates the parametric excitation of the
two-soliton molecule. Ordinary differential equations
(\ref{maineq1}) and (\ref{maineq2}) are solved by fourth-order
Runge-Kutta method. Numerical solution of the GPE (\ref{gpe5}) is
performed by the split-step fast Fourier transform
method.\cite{Press}

\newpage

If the coefficient of nonlocal nonlinearity
$\gamma(t)=2d^2\rho(t)$ of a system vary periodically with time
$\rho(t)=1+\varepsilon\sin(\omega_{0}t+\varphi_{0})$, then an
equilibrium position can be unstable, even if it stable for each
fixed value of the parameter. This instability is what makes it
possible for the parametric excitation of the two-siliton molecule
to appear. The phase $\varphi_{0}$ of the parametric oscillations
undergoes a jump of $-\pi$ as $\omega_{0}$ passes through the
resonance frequency $\omega_{2}$ (see Eq. (\ref{fq2})). When
$\omega_{0}$ is near $\omega_{2}$, beats are observed (Fig.
\ref{fig2}), i. e., the amplitude of the system alternately waxes
when the relation of the phases of the system and the strength of
dipolar interactions are such that the strength of dipolar
interactions rocks the system, communicating energy to it and
wanes when the relation between the phases changes in such a way
that the strength of dipolar interactions brakes the system. The
closer the frequencies $\omega_{0}$ and $\omega_{2}$, the more
slowly the phase relation changes and the large the period of the
beats. As $\omega_{0}\rightarrow\omega_{2}$, the period of the
beats approaches infinity. Since the resonant frequencies are
different for the center-of-mass position $(\omega_{2})$ and width
$(\omega_{1}$, see Eq. (\ref{fq1})$)$ of the two-soliton molecule,
periodic modulation of the parameter $\gamma(t)$ with the
frequency $\omega_{2}$ does not induce resonant oscillations of
the width. The jump in width (see the right panel in Fig.
{\ref{fig2}}) can be explained by asymmetric deformation of the
two-soliton molecule at the impact with the reflecting potential.

\section{Small Amplitude Dynamics}

At the present time, problems of nonlinear oscillations have
attracted much attention in various spheres of physics. In what
follows, we shall consider the dynamics of small nonlinear
oscillations of a system about a position of stable equilibrium.
Near a position of stable equilibrium, a system executes nonlinear
oscillations. It is perfectly natural that the oscillating systems
most accessible for investigations are those with small
nonlinearity because the methods of the theory of perturbations
may be applied to them in some form or the other. Although such an
approximation is entirely legitimate when the amplitude of the
oscillations is sufficiently small, in higher approximations
(called anharmonic oscillations), some minor but qualitatively
different properties of the motion appear.\cite{Bogoliubov,Stoker}
Let us examine in detail the case of two 1D systems of
eigenfrequencies $\omega_{1}$ and $\omega_{2}$ coupled by an
interaction term.

Expanding the coupled system of equations (\ref{maineq1}) and
(\ref{maineq2}) of two variables into a Taylor's series in the
neighborhood of a point $(a_{0},\,\xi_{0})$, and putting for
brevity $x_{1}=a-a_{0}$ and $x_{2}=\xi-\xi_{0}$, we have the
following equations of motion:
\begin{equation}
\ddot{x}_{1}+\omega_{1}^{2}x_{1}=\alpha x_{2}^2, \label{neq1}
\end{equation}
\begin{equation}
\ddot{x}_{2}+\omega_{2}^{2}x_{2}=\beta x_{1}x_{2}. \label{neq2}
\end{equation}
Here, the notation is as follows:
\begin{eqnarray}
\omega_{1} & = &
\bigg\{\frac{3}{a_{0}^4}+\frac{16V_{0}\xi_{0}^2}{\sqrt{\pi}a_{0}^5}
\Big[1-\frac{3}{2}(\xi_{0}/a_{0})^2+\frac{1}{3}(\xi_{0}/a_{0})^4\Big]
e^{-(\xi_{0}/a_{0})^2}+\frac{q N}{2\sqrt{2\pi}a_{0}^3} \nonumber \\
& - & \frac{\gamma\,N(2a_{0}^6-5a_{0}^4w^2+24a_{0}^2w^4-4w^6)}
{4\sqrt{2\pi} (a_{0}^2+w^2)^{9/2}}\bigg\}^{1/2}, \label{fq1}
\end{eqnarray}

\begin{eqnarray}
\omega_{2}=2\big(V_{0}/(\sqrt{\pi}a_{0}^3)\big)^{1/2}\Big[2(\xi_{0}/a_{0})^4
-5(\xi_{0}/a_{0})^2+1\Big]^{1/2}e^{-\frac{1}{2}(\xi_{0}/a_{0})^2}.
\label{fq2}
\end{eqnarray}
The frequencies of small amplitude oscillations remain constant
during the movement of the system. The system under consideration
is weakly coupled. The connection between the frequencies
characterizing the state of the system is determined by the
approximation $\omega_{1}\approx2\omega_{2}$. According to the
expansion of the coupled system of equations (\ref{maineq1}) and
(\ref{maineq2}), the coupling coefficients $\alpha$ and $\beta$
are related by the approximation $\beta\approx2\alpha$. Therefore,
using this integrable case, we can obtain meaningful information
about the motion of the system by considering the integrable
problem in a first approximation.

Stable equilibrium corresponds to a position of the system in
which its potential energy is a minimum. We shall measure the
potential energy from its minimum value. The solution of Eqs.
(\ref{neq1}) and (\ref{neq2}) may be sought in the form
\begin{equation}
x_{1}(t)=A_{1}(t)e^{i\omega_{1}t}+A_{1}^{\ast}(t)e^{-i\omega_{1}t},
\label{sneq1}
\end{equation}
\begin{equation}
x_{2}(t)=A_{2}(t)e^{i\omega_{2}t}+A_{2}^{\ast}(t)e^{-i\omega_{2}t},
\label{sneq2}
\end{equation}
where $A_{1}(t)$ and $A_{2}(t)$ are amplitudes of time which vary
slowly in comparison with the exponential factors, assuming that
$|\ddot{A}_{1}|<<\omega_{1}|\dot{A}_{1}|<<\omega_{1}^2|A_{1}|,
\;|\ddot{A}_{2}|<<\omega_{2}|\dot{A}_{2}|<<\omega_{2}^2|A_{2}|$.
This form of solution is, of course, not exact.

Retaining only the terms with $e^{i\omega_{1}t}$ (corresponding
$e^{i\omega_{2}t}$) and omitting the $|\ddot{A}_{1}|,
\;|\ddot{A}_{2}|$, we have
\begin{equation}
4\omega_{2}\dot{A}_{1}-i\alpha A_{2}^2=0, \label{ameq1}
\end{equation}
\begin{equation}
\omega_{2}\dot{A}_{2}-i\alpha A_{1}A_{2}^{\ast}=0. \label{ameq2}
\end{equation}
One sees easily from the coupled system of equations for the
amplitudes (\ref{ameq1}) and (\ref{ameq2}) that
\begin{equation}
|A_{2}|^2+4|A_{1}|^2=C_{1}=\mathrm{const}, \label{laweng1}
\end{equation}
\begin{equation}
A_{1}A_{2}^{\ast2}+A_{1}^{\ast}A_{2}^2=D_{1}=\mathrm{const}.
\label{const}
\end{equation}

The result obtained signify that the quantity (\ref{laweng1}) is
the law of conservation of energy. Now, using Eqs. (\ref{ameq1})
and (\ref{ameq2}), one obtains
\begin{equation}
\omega_{2}\frac{d}{dt}|A_{2}|^2=i\alpha(A_{1}A_{2}^{\ast2}-
A_{1}^{\ast}A_{2}^2). \label{laweng2}
\end{equation}
Squaring Eq. (\ref{laweng2}) and taking into account Eqs.
(\ref{laweng1}) and (\ref{const}), we get
\begin{eqnarray}
\Big(\frac{d}{dt}|A_{2}|^2\Big)^2 & = &
-\frac{\alpha^2}{\omega_{2}^2}\big[(A_{1}A_{2}^{\ast2}+A_{1}^{\ast}
A_{2}^2)^2-4|A_{1}|^2|A_{2}|^4\big] \nonumber \\
& = & \frac{\alpha^2}{\omega_{2}^2}\big[(C_{1}-|A_{2}|^2)
|A_{2}|^4-D_{1}^2\big]. \label{laweng3}
\end{eqnarray}

Equation (\ref{laweng3}) is similar to the law of conservation of
energy. This equation is similar to equation of motion for a unit
mass particle in the anharmonic potential
\begin{equation}
U(|A_{2}|^2)=(|A_{2}|^2-C_{1})|A_{2}|^4. \label{pot3}
\end{equation}
Here, the parameter $|A_{2}|^2$ appears as a coordinate.

\begin{figure}[htb]
\centerline{\includegraphics[width=10cm,keepaspectratio,clip]{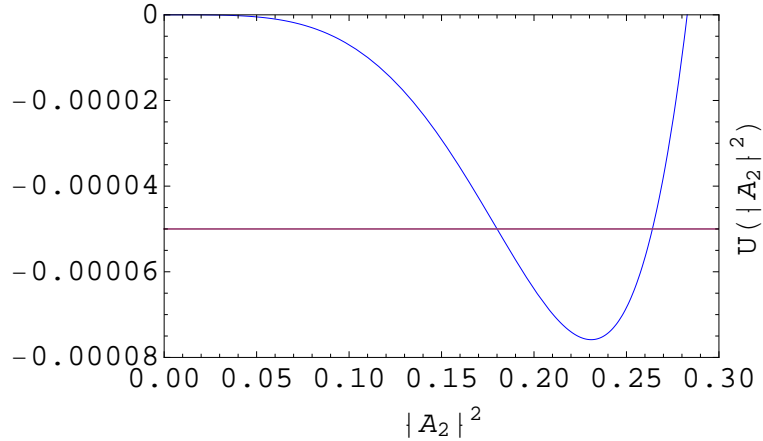}}
\caption{(Color online) The shape of the anharmonic potential
$U(|A_{2}|^2)$ for a given value of the total energy.}
\label{fig3}
\end{figure}

Figure \ref{fig3} illustrates the shape of the anharmonic
potential $U(|A_{2}|^2)$. In this figure, a horizontal line
corresponds to a given value of the total energy $-D_{1}^2$. In
Fig. \ref{fig3}, the movement can only occur in the cavity. The
points at which the potential energy equals the total energy, i.
e. $U(|A_{2}|^2)=-D_{1}^2$ give the limits of the motion.

A finite motion in one dimension is oscillatory, the particle
moving repeatedly back and forth between two points. According to
the reversibility property, the time during which the particle
passes from $|A_{2,1}|^2$ to $|A_{2,2}|^2$ and back is twice the
time from $|A_{2,1}|^2$ to $|A_{2,2}|^2$. Thus, the amplitude
$|A_{2}|$ oscillates, i.e. beats occur. The dependence of the
amplitudes $|A_{1}|$ and $|A_{2}|$ on time can be expressed in
terms of elliptic functions.

Note that in contrast to oscillations of linearly coupled
oscillators in this case, not only the beat depth but also the
period depend on initial amplitudes and phases.

This problem relates to the relationship of longitudinal and
transverse oscillations of a $\mathrm{CO_{2}}$ molecule (so-called
Fermi resonance).\cite{Fermi} The problem under consideration also
occurs in nonlinear optics when the frequency of light doubles and
divides.\cite{Bloembergen}

\section{Action-Angle Variables}

The Hamiltonian formulation of dynamical equations of physical
systems of different nature had a deep impact on the study of
dynamical systems. In this section, we shortly recall the
``action-angle" variables for the integrable Hamiltonian system.
Let us consider the canonical formalism of Hamiltonian system near
stationary points. The Hamiltonian of the coupled system of
equations (\ref{neq1}) and (\ref{neq2}) is
\begin{equation}
H=\frac{\dot{x}_{1}^2}{2}+\frac{\dot{x}_{2}^2}{2}+2\omega_{2}^2x_{1}^2+
\frac{\omega_{2}^2x_{2}^2}{2}-\alpha x_{1}x_{2}^2. \label{h1}
\end{equation}
Introducing the canonical variables $x_{i}$ and
$p_{i}\equiv\dot{x}_{i}\,(i=1,2)$, Eq. (\ref{h1}) is reduced to
the Hamiltonian
\begin{equation}
H=\frac{1}{2}\big(p_{1}^2+p_{2}^2+4\omega_{2}^2x_{1}^2+\omega_{2}^2x_{2}^2
\big)-\alpha x_{1}x_{2}^2. \label{h2}
\end{equation}
In action-angle variables $(I_{i},\theta_{i})$ introduced
as\cite{Landau}
\begin{equation}
x_{i}=\sqrt{2I_{i}/\omega_{i}}\sin\theta_{i},\quad
p_{i}=\sqrt{2I_{i}\omega_{i}}\cos\theta_{i}, \label{a-a1}
\end{equation}
the Hamiltonian can be rewritten as
\begin{equation}
H=2I_{1}\omega_{2}+I_{2}\omega_{2}-\frac{2\alpha
I_{2}}{\omega_{2}}\sqrt{\frac{I_{1}}{\omega_{2}}}\,
\sin\theta_{1}\sin^{2}\theta_{2}. \label{h3}
\end{equation}
Hamilton's equations $\dot{\theta}_{i}=\partial H/\partial
I_{i},\;\dot{I}_{i}=-\partial H/\partial\theta_{i}$ for
action-angle variables $(I_{i},\theta_{i})$ yield the following
two pairs of coupled equations:
\begin{eqnarray}
\dot{I}_{1} & = & \frac{2\alpha
I_{2}}{\omega_{2}}\sqrt{\frac{I_{1}}{\omega_{2}}}\,\sin^{2}\theta_{2}
\cos\theta_{1}, \\
\dot{\theta}_{1} & = & 2\omega_{2}-\frac{\alpha
I_{2}}{\omega_{2}\sqrt{\omega_{2}I_{1}}}\,\sin^{2}\theta_{2}\sin\theta_{1}
\label{a-a2}
\end{eqnarray}
and
\begin{eqnarray}
\dot{I}_{2} & = & \frac{2\alpha I_{2}}{\omega_{2}}
\sqrt{\frac{I_{1}} {\omega_{2}}}\,\sin\theta_{1}\sin2\theta_{2}, \\
\dot{\theta}_{2} & = & \omega_{2}-\frac{2\alpha}{\omega_{2}}\sqrt
{\frac{I_{1}}{\omega_{2}}}\,\sin \theta_{1}\sin^{2}\theta_{2}.
\label{a-a3}
\end{eqnarray}

\begin{figure}[htb]
\centerline{\includegraphics[width=6cm,height=5cm]{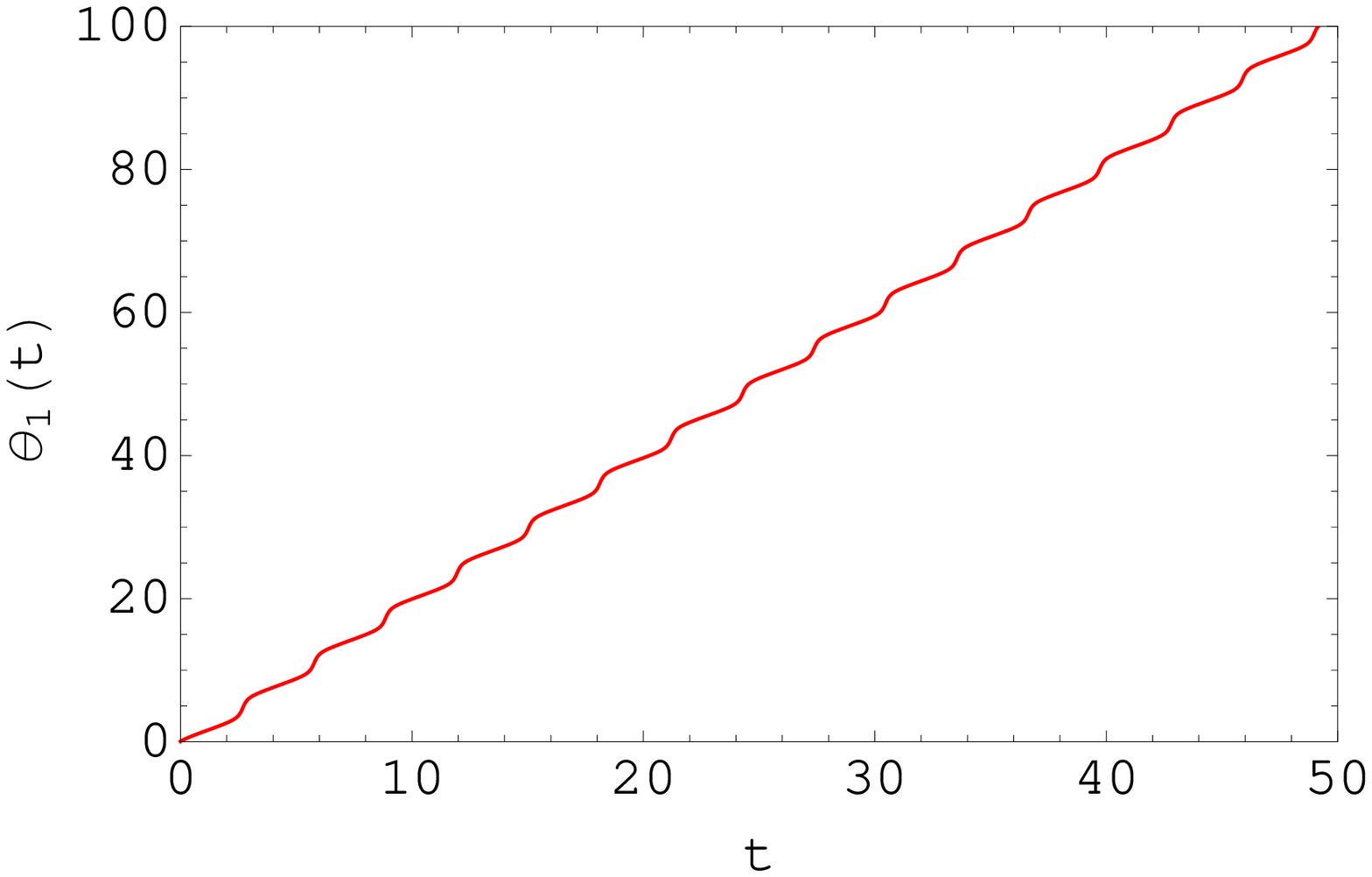}\quad
            \includegraphics[width=6cm,height=5cm]{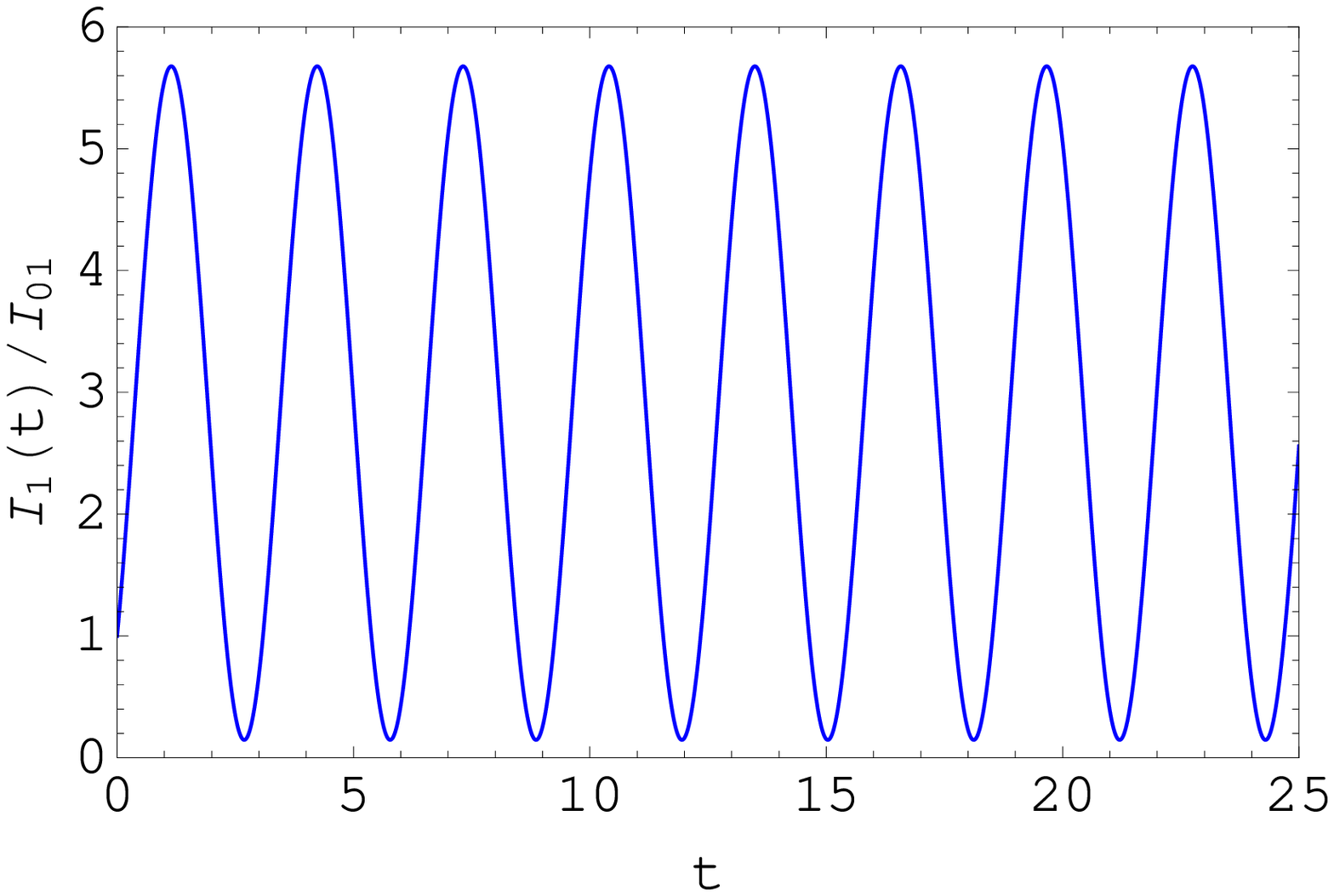}}
\caption{(Color online) Left panel: Evolution of the angle
variable $\theta_{1}(t)$ (red solid line). Right panel:
Oscillations of the action variable $I_{1}(t)$ (blue solid line)
normalized to $I_{01}$.} \label{fig4}
\end{figure}

Let us now consider the stationary points. When the action is
$\dot{I}_{1}=0$, then the stationary value of angle is
$\theta_{01}=\pi/2$. Thus, the stationary value of action is
$I_{01}=\big(\alpha
I_{2}\sin^{2}\theta_{2}/\big(2\omega_{2}^{5/2}\big)\big)^2$ at
fixed values $I_{2},\theta_{2}$. In the second pair of coupled
equations (46) and (\ref{a-a3}), when $\dot{I}_{2}=0$, we obtain
$\theta_{02}=0$ and
$\theta_{02}=\arcsin\big[\big(\omega_{2}^2\sqrt{\omega_{2}/I_{1}}/(2\alpha)
\big)^{1/2}\big]$ at fixed value $I_{1}$. Let us find the
numerical solutions of the first pair of coupled equations (44)
and (\ref{a-a2}) for the action $I_{1}(t)$ and angle
$\theta_{1}(t)$ variables of Hamiltonian system. In Fig.
\ref{fig4}, we represent the evolution of the angle variable and
oscillations of the action.

\newpage

The formulation of Hamiltonian equations in action-angle variables
is most convenient to study Hamiltonian systems in the presence of
perturbations and to construct symplectic maps.\cite{Zaslavsky}
The difficulty of quantizing non-integrable systems was expressed
in terms of the invariant tori of action-angle variables. The use
of action-angle variables was central to the solution of the Toda
lattice,\cite{Toda} and to the definition of Lax pairs.\cite{Lax}

\section{Stationary State of a Two-Soliton Molecule}

Below, we consider the stationary width of a two-soliton molecule.
Equation (\ref{maineq1}), which allows us to find the stationary
solution of the one-dimensionless nonlocal GPE (\ref{gpe5}) and
describes its dynamics near the fixed point, is the main result of
this section. At large deviations from the stationary state, the
waveform (\ref{ansatz1}) can be deviated from the Gauss-Hermite
shape, and the predictions of the variational approximation become
less accurate.

The stationary width of a two-soliton molecule $a_{0}$ is
calculated from the following condition:
\begin{eqnarray}
\frac{a_{0}^4(a_{0}^4+4w^4)}{(a_{0}^2+w^2)^{7/2}}-\frac{q}{\gamma}\,a_{0}-
\frac{16\sqrt{2}V_{0}\xi_{0}^2}{\gamma\,N
a_{0}}\Big(1-\frac{2\xi_{0}^2}{3a_{0}^2}\Big)e^{-(\xi_{0}/a_{0})^2}-\frac
{4\sqrt{2\pi}}{\gamma\,N}=0. \label{stwd}
\end{eqnarray}

\begin{figure}[htb]
\centerline{\includegraphics[width=10cm,keepaspectratio,clip]{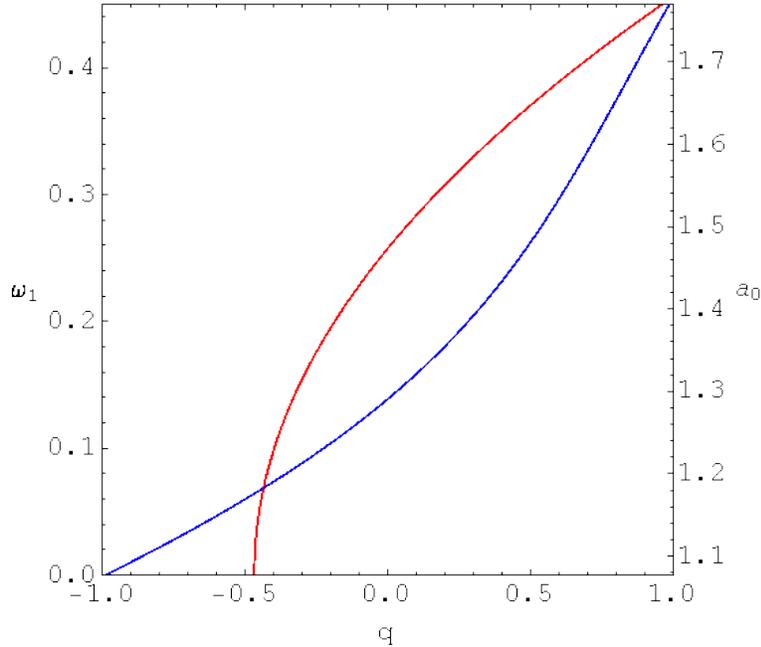}}
\caption{(Color online) The frequency of low energy shape
oscillations $(\omega_{1})$ of a two-soliton molecule in a dipolar
BEC as a function of the strength of contact interactions $q$,
according to Eq. (\ref{fq1}) (red solid line), and the stationary
width of a two-soliton molecule ($a_{0})$ obtained from Eq.
(\ref{stwd}) (blue line). Parameter values: $V_{0}=1,\,N=4,\,w=5$
and $\gamma=20$.} \label{fig5}
\end{figure}

Figure \ref{fig5} illustrates the frequency of small amplitude
oscillations of the two-soliton molecule as a function of the
strength of contact interactions, according to Eq. (\ref{fq1}).
The same figure shows the stationary width of the two-soliton
molecule for a given strength of contact interactions, obtained
from Eq. (\ref{stwd}).

\newpage

The case of repulsive contact interactions $(q<0)$ give rise to
decreasing of the frequency of oscillations compared to the case
of pure dipolar interactions $(q=0)$, while the contribution of
attractive interactions $(q>0)$ is opposite, leading to increasing
of the oscillations frequency.

Low energy collective oscillations of ultracold atoms can provide
essential information about the interatomic forces in
BEC.\cite{Jin} As a result, the expression (\ref{fq1}) for the
frequency of the small amplitude oscillations of the two-soliton
molecule can be useful in experiments with ultracold polar
molecules.

In nonlinear wave equations, the stability of localized solutions
can be examined by means of the Vakhitov-Kolokolov
criterion.\cite{Vakhitov-Kolokolov} Using the usual
procedure,\cite{Abdullaev2} we look for the stationary solution of
the one-dimensionless nonlocal GPE (\ref{gpe5}) in the form
$\psi(x,t)=e^{-i\mu t}\zeta(x)$, with $\mu$ denoting the chemical
potential.

The time-independent one-dimensionless nonlocal GPE (\ref{gpe5})
takes the form
\begin{equation}
\big(\mu-kx-V(x)\big)\zeta+\frac{1}{2}\,\zeta_{xx}-q\,\zeta^3+\gamma\zeta\int
\limits_{-\infty}^{\infty}R(|x-z^{\prime}|)\zeta^{2}(z^{\prime})dz^{\prime}=0.
\label{gpe6}
\end{equation}
The corresponding Lagrangian density is
\begin{equation}
{\cal
L}=\frac{1}{4}\Big[\zeta_{x}^{2}-2\big(\mu-kx-V(x)\big)\zeta^{2}+
q\,\zeta^{4}-\gamma\zeta^{2}\int\limits_{-\infty}^{\infty}
R(|x-z^{\prime}|)\zeta^{2}(z^{\prime})dz^{\prime}\Big].
\label{lagdensity3}
\end{equation}
We seek a solution in the form:
\begin{equation}
\zeta(x)=A(x-\xi)\exp\Big[{-\frac{(x-\xi)^2}{2a^2}}\Big].
\label{ansatz2}
\end{equation}
Substituting the trial function (\ref{ansatz2}) and using the
response function (\ref{ker}) into the Lagrangian density
(\ref{lagdensity3}) and subsequently integrating over the space
variable, we obtain the following averaged Lagrangian
\begin{eqnarray}
L & = & \frac{3\sqrt{\pi}A^2a}{16}-\frac{\sqrt{\pi}\mu
A^2a^3}{4}+\frac{\sqrt{\pi}k
A^2a^3\xi}{4}+\frac{V_{0}A^2\xi^2}{2}\,e^{-(\xi/a)^2} \nonumber \\
& + & \frac{3\sqrt{\pi/2}\,q
A^4a^5}{64}-\frac{\sqrt{\pi/2}\,\gamma
A^4a^6}{64}\bigg[\frac{4w^2(a^2+w^2)+3a^4}{(a^2+w^2)^{5/2}}\bigg].
\label{lagrangian4}
\end{eqnarray}
Performing further the standard variational approximation
procedure,\cite{Anderson,Malomed} we obtain the following
functions of the variable $a$ for the chemical potential and norm:
\begin{eqnarray}
\mu(a) & = &
\frac{1}{2a^2}+k\,\xi_{0}+\frac{V_{0}\xi_{0}^2}{\sqrt{\pi}a^3}
\Big(1+\frac{2\xi_{0}^2}{3a^2}\Big)e^{-(\xi_{0}/a)^2}+\frac{11q\,N}
{16\sqrt{2\pi}a} \nonumber \\
& - & \frac{\gamma\,N}{16\sqrt{2\pi}}\bigg[\frac{11a^6+
28a^2w^2(a^2+w^2)+16w^6}{(a^2+w^2)^{7/2}}\bigg], \label{chempot}
\end{eqnarray}

\begin{eqnarray}
N(a)=\frac{8\big[3\sqrt{\pi}a^3+4V_{0}\xi_{0}^2(3a^2-2\xi_{0}^2)
e^{-(\xi_{0}/a)^2}(a^2+w^2)^{7/2}\big]}{3\sqrt{2}a^4\big[\gamma
\,a^3(a^4+4w^4)-q(a^2+w^2)^{7/2}\big]}. \label{norm}
\end{eqnarray}
We have established a connection between the chemical potential
and the norm.\cite{Pethick} According to Vakhitov-Kolokolov
criterion\cite{Vakhitov-Kolokolov} $\frac{d\mu}{dN}<0$, we
obtained the stability of the two-soliton molecule. It should be
noted that for the repulsive contact interactions $(q<0)$, pure
dipolar interactions $(q=0)$ and the effect of attractive
interactions $(q>0)$, the two-soliton molecule remains stable for
different values of the nonlocal coefficient $\gamma\,(\gamma>0)$.
The stronger attraction between solitons leads to more stability
of the molecule.\cite{Lahaye}

\section{Conclusion}

The model of a ``quantum bouncer" has been extended to a dipolar
BECs. We have studied the effects of atomic dipole-dipole
interactions and gravity on the dynamics of BECs by means of
variational approximation and numerical simulations. In numerical
experiments, we observed the parametric excitation by a
two-soliton molecule when the vertical position of the atomic
mirror is periodically varied in time. We have provided thorough
comparison between the results of variational approximation and
numerical simulations of the GPE. We have studied Hamilton's
dynamic system for a dipolar condensates in terms of
``action-angle" variables. In this paper, the stationary state of
a two-soliton molecule in a dipolar BECs has been studied.

\section*{Acknowledgements}

The author would like to thank the workshop of the Laboratory of
Theoretical Physics of the Physical-Technical Institute. This work
has been supported by Grant No. FA-F2-004 of the Agency for
Science and Technologies of Uzbekistan. \\

\end{document}